\documentclass[preprintnumbers,amsmath,amssymb,11pt]{revtex4}
\usepackage{graphicx}
\usepackage{dcolumn}
\usepackage{bm}
\usepackage[dvips,colorlinks,bookmarksopen,bookmarksnumbered,citecolor=red,urlcolor=red]{hyperref}

\pdfoutput=1
\begin{document}

\title{Two-dimensional Moist Stratified Turbulence and the Emergence of Vertically Sheared Horizontal Flows}
\author
{Jai Sukhatme$^1$, Andrew J. Majda$^2$ and Leslie M.\ Smith$^{3,4}$ \\
   1. Centre for Atmospheric and Oceanic Sciences, Indian Institute of Science, Bangalore 560012, India \\
   2. Courant Institute of Mathematical Sciences, New York University, New York, NY 10012 \\
   3. Mathematics Department, University of Wisconsin-Madison, Madison, WI 53706 \\
   4. Engineering Physics Department, University of Wisconsin-Madison, Madison, WI 53706 \\}
\date{\today}
\begin{abstract}

Moist stratified turbulence is studied in a two-dimensional 
Boussinesq system influenced by
condensation and evaporation. 
The problem is set in a periodic domain and employs simple evaporation and condensation schemes, wherein both the processes push parcels towards saturation. 
Numerical simulations
demonstrate the emergence of a moist turbulent state consisting of
ordered structures with a clear power-law type spectral scaling from
initially spatially uncorrelated conditions.
An asymptotic analysis in the limit of
rapid condensation and strong stratification
shows that, for initial conditions with enough water substance to saturate the domain, the equations support a straightforward
state of moist balance characterized by a hydrostatic, saturated, vertically sheared horizontal flow (VSHF).
For such initial conditions, by means of long time numerical simulations, the emergence of moist balance is verified. Specifically,
starting from uncorrelated data, subsequent to the development of a moist turbulent state,
the system experiences a rather abrupt transition to a regime which is close to saturation and dominated
by a strong VSHF. 
On the other hand, initial conditions which do not have enough water substance to saturate the domain, do not attain moist balance. Rather, the 
system remains in a turbulent state and oscillates about moist balance.
Even though balance is not achieved with these general initial conditions, the time scale of oscillation about moist balance is much larger than 
the imposed time scale of condensation and
evaporation, thus indicating a distinct dominant slow component in the moist stratified two-dimensional turbulent system.

\end{abstract}

\maketitle

\section{Introduction}

Water, in its different phases, plays a significant role in shaping climate of the Earth \cite{Ray-nature},\cite{Sch-review}.
From a dynamical perspective, the energy release associated with
changes in phase of water substance has a profound effect on specific atmospheric phenomena
as well as the large-scale general
circulation of the tropical and extra-tropical troposphere \cite{ENB},\cite{Frier}. 
To develop an understanding of the interaction of advection and condensation/evaporation, 
we study a moist stratified two-dimensional (2D) system which consists of 
nonlinear gravity wave-modes (including zero frequency vertically sheared horizontal flows - VSHF),
and a dynamically consistent heat source due to condensation/evaporation
(note that the kinematic advection condensation problem, where the advecting flow is prescribed is also of significant
practical interest \cite{SY}). 
The decomposition of the stratified Boussinesq equations into 
vortical and wave modes is specified in Lelong and Riley \cite{LR} (applications can be 
found in Bartello \cite{Bartello}, Sukhatme \& Smith \cite{SS-gafd}), mathematical details and a description of the VSHF can be found
in Embid \& Majda \cite{Embid1998} and Smith \& Waleffe \cite{SW-jfm}.

The stably stratified 2D Boussinesq equations (without the influence of water substance) have been studied
as a model of thermal
convection \cite{CMV}. Numerical work on the scaling
of kinetic and potential energy in this system is seen to be quite non-universal : in addition to
the original Obhukov-Bogliano
postulates, scaling laws for the temperature variance (or entropy) have been proposed and realized
\cite{Lvov},\cite{Brand},\cite{Toh}. Unlike pure 2D turbulence, the stably stratified system does not
conserve enstrophy. In spite of this simulations suggest the possibility of an
inverse transfer of kinetic
energy, the formation of a VSHF and a forward cascade of temperature variance (or potential energy) \cite{Brand},\cite{Smith-short},\cite{Toh}.
On the other hand, noting that the 2D system involves the nonlinear interaction of gravity waves, the
susceptibility of these waves to parametric subharmonic instability leads to the anticipation of a robust
transfer of
energy to smaller scales \cite{McEwan-Rob},\cite{Mied},\cite{Klos}. Numerical work in a geophysical
context has documented this forward transfer of energy
\cite{Orlanski},\cite{BS},\cite{BS1},\cite{SS}. Further, the possible development of singularities
\cite{Pumir} and fronts \cite{E},\cite{SS} as well as the
global well-posedness of these equations has also attracted significant attention
\cite{MB-book},\cite{CF},\cite{Hmidi},\cite{Hou}. 
One of the main issues we investigate in the present work is whether 
the moist stratified two-dimensional system is capable of exploiting 
the available potential energy in a systematic manner so as to spontaneously organize 
spatially uncorrelated initial data.

Two-dimensional stratified models have been used quite extensively in atmospheric settings, for example the tendency to organize condensing 
convection by gravity waves in 2D has been noted 
by Huang \cite{Huang}. Similarly
moist 2D models have been used for studying tropical cyclone formation \cite{MoFa},\cite{MMX2}, squall-line thunderstorms \cite{Seitter-Kuo},\cite{MX}, 
atmospheric density currents \cite{Seitter},
unstable modes in moist convection \cite{Asai} as well as in elucidating properties of precipitating 
convection \cite{Emanuel-precip}. In another vein, 2D models arise in the so-called super-parameterization schemes for 
general circulation modeling wherein 
a high resolution 2D model is embedded in a coarser three-dimensional (3D) model to help resolve features of geophysical systems which 
support an extended range of scales \cite{Grab-S},\cite{Grab},\cite{Majda-super}. 
Quite naturally, as the 2D model feeds into the larger scale 3D flow, it is important to 
develop a feel for 2D moist turbulence in its own right. 

In addition to the scaling and structure formation, we hope to address the issue of balance and "slowness" in this simplified moist system. 
Indeed, the notion of balance and the existence (or not) of a slow-manifold in geophysical systems has a rich history
\cite{Lorenz},\cite{Leith},\cite{Warn}. We refer the reader to the text by Daley \cite{Daley} for an introduction and to 
Vanneste \cite{Vanneste} (and the references therein) for recent progress in the subject. Loosely, given the wide 
range of temporal scales inherent in geophysical systems, it is natural to inquire whether physically 
relevant information is actually contained in an independent simpler reduced dynamical system that evolves on a comparatively slower time scale. 
Usually such a simplified or reduced description is possible in certain limiting scenarios, the celebrated example being the emergence of quasigeostrophic
dynamics in the limit of small Froude and Rossby numbers for a rotating and stratified fluid \cite{Embid1998},\cite{Babin}.
In particular, for the system at hand, by 
imposing strong stratification (which implies the presence of high frequency gravity waves) and rapid condensation/evaporation (which 
implies rapid heating/cooling) we would like to see if it is possible to asymptotically extract an emergent state of balance or a reduced system 
which is able to capture the main features of the entire system. If in fact the analysis points to such a simplified description, it is also
of interest to examine whether general initial conditions are capable of attaining this state. 

We begin by introducing the basic equations and the schemes employed for condensation and evaporation. 
The physical problem at hand is to consider the fate of perturbations to a basic state which is saturated and in 
hydrostatic balance. A linear analysis in a periodic domain reveals that the most unstable modes have a columnar structure whose growth rate has a 
particularly simple form in the limit of rapid condensation (and evaporation). The predicted emergence of columnar structures with appropriate 
growth rates is then realized in linear and non-dissipative simulations. 
Another aspect of the linear analysis is the presence of a distinct 
slow mode, this mode is also obtained via an asymptotic expansion of the suitably nondimensionalized nonlinear equations. The slow mode represents 
a state of moist balance and is
characterized by a hydrostatic, saturated VSHF. 
To test the accessibility of this state of balance as well as scaling and structure formation in this moist system we then proceed to
nonlinear simulations under conditions of 
strong stratification and rapid condensation/evaporation.
Employing spatially uncorrelated initial conditions, we 
examine the development of a turbulent state which includes the formation of coherent structures and the spectral scaling of the vapor, liquid, velocity
and temperature  
fields. After the emergence of a turbulent state, 
initial conditions which
have sufficient total water content to saturate the domain 
are seen to abruptly 
transition to the aforementioned state of moist balance.  
For initial conditions which are insufficient in the above mentioned sense, the flow exhibits similar scaling and structure formation. However, due to
the lack of sufficient water substance, the
system cannot achieve moist balance. Rather, our numerical examination shows that the system remains in a turbulent state and oscillates in an aperiodic or quasiperiodic manner 
about the state of moist balance.
Finally, a summary and discussion conclude the paper. 

\section {Governing Equations}

The moist Boussinesq framework has been the subject of numerous recent investigations : for example, in the scaling of 
moist turbulence \cite{Spy}, 
the modelling of precipitation fronts \cite{FPM}, moist Rayleigh Bernard convection \cite{PS}, studying droplet growth in a 2D 
setting \cite{CMST}, tropical cyclone formation \cite{MMX2} and squall-line thunderstorms \cite{MX}. The governing equations are : 

\begin{eqnarray}
\frac{D \vec{u}}{D t} = - \frac{\nabla p'}{\rho_0} + (\frac{g}{T_0}) T' \hat{k}, \nonumber \\
\frac{D T'}{D t} = - \lambda w + \frac{\mathcal{L}}{C_p} S, \nonumber \\
\frac{D q_v}{D t} = -S \nonumber ,~
\frac{D q_l}{D t} = S, \nonumber \\
\frac{\partial u}{\partial x} + \frac{\partial w}{\partial z} = 0.
\label{1}
\end{eqnarray}
In the above $\vec{u}=(u,w)$ is the 2D incompressible velocity field while $T',p'$ are the potential temperature and pressure perturbations.
$\mathcal{L}$ is the latent heat of condensation and $C_p$ is the specific heat.
We consider an equation of state for an ideal gas wherein $p=\rho R T$.
In effect these equations
isolate the effect of condensation (evaporation) via dynamical heating (cooling). The heating and cooling are represented by the source $S$ which is described later in the text. 
We have not included a rain-out velocity or the influence of water substance on the buoyancy and the equation of state. 
A discussion of these effects and their role in the moist 
framework can be found in Klein \& Majda \cite{KM}. 

As per the usual Boussinesq setup, the 
buoyancy frequency is given by $(g\lambda/T_0)^{\frac{1}{2}}$ where $g,T_0,\lambda$ denote the gravitational acceleration, 
reference potential temperature and the background potential temperature gradient, respectively (i.e.\ $T=T_0 + \lambda z + T'$).
In addition, 
$q_v(x,z,t)$ and $q_l(x,z,t)$ are the mixing ratios of water vapor and liquid water, and
$q_s(x,z)$ is a fixed saturation mixing ratio profile. In essence, the 
mixing ratios are governed by advection-condensation-evaporation equations, and $q=q_v+q_l$ (the total 
water content) is materially conserved, i.e. $Dq/Dt=0$. 

Physically, ${q_v}$ is advected
via the fluid flow and if it exceeds the saturation value at the new destination, the mixing ratio
proceeds to
relax back to its saturation value. Similarly ${q_l}$ is advected via the flow,
and whenever ${q_l} > 0$ and
${q_v} < q_s$, the parcel experiences evaporation.
These statements are quantified via 

\begin{eqnarray}
C =  \frac{1}{\tau_c} ({q_v}(\vec{x},t) - q_s(\vec{x})) ~H({q_v}(\vec{x},t) - q_s(\vec{x})) ,~
E = \frac{1}{\tau_e} {q_l} ~H(q_s(\vec{x}) - {q_v}(\vec{x},t)).
\label{1b}
\end{eqnarray}
$E$ and $C$ represent evaporation and condensation, and $\tau_c,\tau_e$ are the time scales associated with these two processes respectively. 
$H(\cdot)$ denotes the Heaviside step function. Equation (\ref{1b}) encodes an essential simplification
that at any given instant of time a parcel
experiences either condensation or evaporation. 
In the limit of rapid condensation (and evaporation), 
i.e.\ $\tau_c,\tau_e \rightarrow 0$, we have 
$0 \le q_v(\vec{x},t) \le q_s(\vec{x})$ and $0 \le q_l(\vec{x},t) \le q_l(\vec{x}_0,0) + q_s(\vec{x}_0)$ --- where 
$(\vec{x}_0,0)$ is the Lagrangian origin of the parcel at $(\vec{x},t)$. One can see that
the upper bound on $q_v$ is local while that on $q_l$ is spatio-temporally non-local. 
Finally, the source in (\ref{1}) is given by $S=C-E$. 

{\it Conservation properties :} Apart from the material invariance of the total water substance ($Dq/Dt=0$), (\ref{1})
also preserves the material conservation of moist static energy given by $\frac{D}{Dt} (T + \frac{L}{C_p} q_v) = 0$. With regard 
to an energy conservation law, (\ref{1}) with periodic boundaries yields

\begin{equation}
\frac{\partial}{\partial t} \int [u^2 + w^2 + \frac{g}{T_0 \lambda} {T'}^2] = \frac{g}{T_0 \lambda} \frac{\mathcal{L}}{C_p} \int S T'.
\label{1c}
\end{equation}
Therefore, the heating and temperature fluctuations have 
to be correlated for a growth in
perturbation energy \cite{ENB} --- a detailed discussion of the conservation properties of moist systems that include bulk cloud 
microphysics can be found in Frierson, Paulius \& Majda \cite{FPM}. 

{\it Vorticity-stream formulation :} We introduce a streamfunction ($\psi$), with $u=-\partial_z \psi,
w=\partial_x \psi$ and vorticity $\omega = - \triangle \psi$. 
Further we consider $q_s = q_s(z) = q_0 - \beta z$, i.e.\ the saturation mixing ratio is a linear function of $z$. 
Defining ${q_v}'=q_v-q_s$, (\ref{1}) can
be re-written as

\begin{eqnarray}
\frac{D \omega}{D t} = - (\frac{g}{T_0}) \frac{\partial T'}{\partial x}, \nonumber \\
\frac{D T'}{D t} = - \lambda \psi_x + \frac{\mathcal{L}}{C_p} S, \nonumber \\
\frac{D {q_v}'}{D t} = \beta \psi_x - S  ,~
\frac{D q_l}{D t} = S.
\label{2}
\end{eqnarray}

{\it Linear analysis : } Linearizing (\ref{2}) about our base state of rest, hydrostatic balance, no liquid water and saturation 
(i.e.\ $S=0$), we obtain a constant co-efficient system of partial differential equations. Substituting plane wave solutions of 
the form $\exp\{ [{\rm i} \vec{k}\cdot \vec{x} - \sigma t] \}$, in addition to a distinct $\sigma=0$ mode, we get 

\begin{equation}
\sigma^3 - \sigma^2 (\frac{\delta_1}{\tau_c} + \frac{\delta_2}{\tau_e}) + \sigma \frac{k_x^2}{k^2} (\frac{\lambda g}{T_0}) - 
\frac{k_x^2}{k^2} [ \frac{\lambda g}{T_0} (\frac{\delta_1}{\tau_c} + \frac{\delta_2}{\tau_e}) - \frac{\mathcal{L}}{C_p} \frac{\beta g}{T_0}\frac{\delta_1}{\tau_c} ] = 0,
\label{3}
\end{equation}
where $\delta_1=H(q_v-q_s)$ and $\delta_2=H(q_s-q_v)$. As expected, when $(\delta_1,\delta_2)=(1,0)$ i.e.\ during condensation, 
$\frac{\mathcal{L}}{C_p} \beta >
\lambda$ leads to an unstable situation. 
Further, when $\tau_c \rightarrow 0$ (rapid condensation) (\ref{3}) yields
$\max(|\sigma|) = \sqrt{\frac{g \delta}{T_0}}$ where $\delta=\frac{\mathcal{L}}{C_p} \beta - \lambda$ and this maximal growth is attained for
fields independent of $z$. Therefore, columnar or vertically coherent structures are the fastest growing modes in linear rapid 
condensation. 

The slow mode $\sigma =0$ is possible only when $\delta_1=\delta_2=0$, i.e.\ $q_v=q_s$. In addition, this mode satisfies 
hydrostatic balance with $w=0$ implying $u=u(z)$. 
Hence, the slow dynamics of the linear system consists of
a hydrostatically balanced, saturated VSHF.


{\it Nondimensionalization : } Nondimensionalizing the momentum equations and the incompressibility condition from (\ref{1}), we obtain (dropping primes)

\begin{eqnarray}
\frac{D u}{D t} = - (Eu) \frac{\partial \phi}{\partial x}, \nonumber \\
\frac{D w}{D t} = - (Eu) \frac{\partial \phi}{\partial z} + (\frac{\Gamma_\theta}{Fr^2}) T, \nonumber \\
\frac{\partial u}{\partial x} + \frac{\partial w}{\partial z} = 0.
\label{4}
\end{eqnarray}
Here $Eu=\mathcal{P}/(\rho_0 U^2)$ and $Fr=U/(NL)$ are the Euler and Froude numbers, and $\phi=p/\rho_0$. $L,\mathcal{P}$ and $U$ are the scale parameters 
of the length, pressure and velocity fields respectively. 
$\Gamma_\theta$ is a nondimensional number defined by
$\Gamma_\theta=\Theta/(\lambda L)$ where the potential temperature is nondimensionalized using the scale parameter $\Theta$. 

In addition, the energy and water substance equations take the nondimensional form

\begin{eqnarray}
\frac{D T}{D t} = - (\frac{1}{\Gamma_\theta}) w + (\frac{\mathcal{L}Q_0}{C_p \Theta})(\frac{\tau_a}{\tau_c}) [(q_v-q_s) H(q_v-q_s) - q_l H(q_s -q_v)],   \nonumber \\
\frac{D q_v}{D t} = (\frac{\tau_a}{\tau_c}) [-(q_v-q_s) H(q_v-q_s) + q_l H(q_s -q_v)],   \nonumber \\
\frac{D q_l}{D t} = (\frac{\tau_a}{\tau_c}) [(q_v-q_s) H(q_v-q_s) - q_l H(q_s -q_v)],   
\label{4a}
\end{eqnarray}
where $\tau_a= U/L$ represents the advective time scale and $q_v,q_l$ and $q_s$ have been scaled by $Q_0 = \max{[q_s]}$. 
Therefore, taken together (\ref{4}) and 
(\ref{4a}) are the nondimensional forms of the governing equations.

Since our interest is in stratified systems, we set $Fr = \epsilon < 1$ in the following asymptotic analysis.
Further, considering small potential temperature variations to the background state, we set $\Gamma_\theta = \epsilon$ and choose $Eu=1$. 
Clearly, the influence of heating on the dynamics is likely to be significant when the time scale
associated with condensation (and evaporation) is smaller than the advective time scale, i.e.\
$\frac{\tau_a}{\tau_c} = \frac{1}{\epsilon}$. This choice is also consistent with the time scales associated with advection (of the order of days) and condensation 
(of the order of a few hours) in the 
atmosphere \cite{FPM}. Finally, given the large latent heat of water substance we scale $(\mathcal{L} Q_0)/(C_p \Theta) = \frac{1}{\epsilon}$.
With these substitutions, (\ref{4}) and (\ref{4a}) read

\begin{eqnarray}
\frac{D u}{D t} = - \frac{\partial \phi}{\partial x}, \nonumber \\
\frac{D w}{D t} = - \frac{\partial \phi}{\partial z} + \frac{1}{\epsilon} T, \nonumber \\
\frac{D T}{D t} = - \frac{1}{\epsilon} w + \frac{1}{\epsilon^2} [(q_v - q_s) \delta_1 - q_l \delta_2],  \nonumber \\
\frac{\partial u}{\partial x} + \frac{\partial w}{\partial z} = 0, \nonumber \\
\frac{D q_v}{D t} = \frac{1}{\epsilon} [-(q_v - q_s) \delta_1 + q_l \delta_2],  \nonumber \\
\frac{D q_l}{D t} = \frac{1}{\epsilon} [(q_v - q_s) \delta_1 - q_l \delta_2].  
\label{5}
\end{eqnarray}

{\it Asymptotic analysis :} As the linear analysis reveals an instability during condensation, we examine the dominant balances 
when $\delta_1=1$ and $\delta_2=0$.
For small $\epsilon$, plugging in an asymptotic expansion of the form $f = f^0 + \epsilon f^1 +...$ for all fields,
at $O(1/\epsilon^2)$ we obtain

\begin{equation}
q_v^0 = q_s,
\label{6}
\end{equation}
and at $O(1/\epsilon)$ we obtain

\begin{equation}
T^0= 0 ~,~ 0 = -w^0 + q_v^1.
\label{6a}
\end{equation}
Carrying forward to $O(1)$ we have

\begin{eqnarray}
\frac{D^0}{D t} u^0 = - \frac{\partial \phi^0}{\partial x}, \nonumber \\
\frac{D^0}{D t} w^0 =  - \frac{\partial \phi^0}{\partial z} + T^1, \nonumber \\
\frac{\partial u^0}{\partial x} + \frac{\partial w^0}{\partial z} = 0, \nonumber \\
\frac{D^0}{D t} T^0 =  -w^1 + q_v^2, \nonumber \\
\frac{D^0}{D t} q_v^0 =  -q_v^1, \nonumber \\
\frac{D^0}{D t} q_l^0 =  q_v^1. 
\label{6b}
\end{eqnarray}
Here $D^0/Dt = \partial_t + u^0 \partial_x + w^0 \partial_z$, i.e.\ it represents advection via the order zero velocity fields.

Let us make a few observations from equations (\ref{6}), (\ref{6a}) and (\ref{6b}). As $T^0=0$, the fourth equation of (\ref{6b}) gives $w^1=q_v^2$. This establishes
a hierarchical link between the nondimensional vertical velocity and vapor field, specifically $w^0=q_v^1$ and $w^1=q_v^2$. Further, adding the last two equations in (\ref{6b})
gives

\begin{equation}
\frac{D^0}{D t} (q_v^0 + q_l^0) = 0.
\label{6c}
\end{equation}
Therefore, the order zero water substance is materially conserved under evolution via the zero order flow. 
We now specialize to the case wherein $q_s=q_s(z)$, keeping this in mind the vapor equation in (\ref{6b}) gives (using $w^0 = q_v^1$)

\begin{equation}
q_v^1 \frac{\partial q_s}{\partial z} = - q_v^1.
\label{6e}
\end{equation}
If (\ref{6e}) is to hold for general saturation profiles, it implies that $q_v^1 = 0$. But, at the present level of approximation, $q_v = q_v^0 + \epsilon q_v^1$. 
Therefore, $q_v^0=q_s$ and $q_v^1=0$ can be true globally only if the initial water substance is sufficient to saturate the domain.

For such initial conditions, collecting the information in (\ref{6}), (\ref{6a}) and the liquid water equation in (\ref{6b}), we see that 
the vapor field tends to saturation, and the excess vapor is converted to liquid water. Further, as 
$w^0=0$ (when $q_v^1=0$), this implies a hydrostatic balance between $\phi^0$ and $T^1$, and from incompressibility we obtain a 
VSHF, i.e.\ $u^0=u^0(z,t)$. 
With this information, the liquid water evolution in (\ref{6b}) 
simplifies to

\begin{equation}
\frac{\partial}{\partial t} q_l^0 + u^0 \frac{\partial q_l^0}{\partial x} =0.
\label{6f}
\end{equation}
As $q_v \rightarrow q_s$, we have $\delta_1 \rightarrow 0$ and $\delta_2=0$. Hence,  
this is the nonlinear version of the slow mode ($\sigma=0$) encountered in the linear analysis of the problem. 
In fact, following Majda, Mohammadian and Xing \cite{MMX}, we refer to this as a state of moist balance. 
From the preceding analysis, we deduce that any
initial condition with sufficient water substance may evolve, possibly at long times, towards moist balance.
Whether such a state is actually attained is
examined numerically in the latter portion of the present work.

Quite clearly, initial conditions which do not have enough water substance to saturate the domain cannot satisfy $q_v^0=q_s$ and $q_v^1=0$ globally, and
therefore cannot achieve a state of moist balance. In fact, we intuitively expect such initial conditions to evolve around moist balance, 
with a continual interplay of condensation and evaporation in 
different portions of the domain. It is unclear if, under these circumstances, the gross features of the system will evolve at a timescale comparable to $\tau_c$ (and $\tau_e$)
or whether a slower timescale (for example, the advective scale) will come to dominate the dynamics. Indeed, another aim of our numerical work is to quantify the 
long term fate of such initial conditions.

\section{Numerical examination of the system}

Numerically, we solve the system in the vorticity-stream form using a pseudo-spectral method in a periodic domain of size $L_d$. The time stepping 
is via a fourth order Runge Kutta scheme. Dissipation is of the form of a hyperviscous damping given by 
$(-1)^{n+1} \nu_n \triangle^{n} f$ (for all fields $f$). In particular, we adopt the 
formulation employed by Maltrud and Vallis \cite{Maltrud-Vallis} wherein $\nu_n = f_{\textrm{rms}}/({k_m}^{2n-2})$, where
$k_m$ is the maximum resolved wavenumber. In all simulations the hyperviscous order is eight. 

The equations solved are (\ref{2}), and we set $g/T_0=\lambda=\beta=1/\epsilon$ (along with $\mathcal{L}/C_p = 1/\epsilon$, this amounts to a 
solving (\ref{5}) in vorticity stream form). The simulations are carried out for a range of 
$\epsilon$ values that spans $10^{-1}$ to $10^{-4}$ for linear runs and $10^{-1}$ to $10^{-2}$ for the nonlinear runs.
This gives the required leading order dependence of $Fr \sim \Gamma_\theta = \epsilon$.
The linear saturation profile is prescribed as $q_s=Q_0[1-z/L_d]$, where $Q_0=L_d/\epsilon$. As $\beta=\lambda$, the condition for linear instability 
reduces to $\mathcal{L}/C_p > 1$. 
In the atmosphere, for water substance, $\mathcal{L}/C_p \approx 10^3$, therefore along with small $Fr$ and $\Gamma_\theta$, the instability criterion 
is satisfied.
In the nonlinear simulations we set this ratio to $1/\epsilon$. 
For the linear runs, we have used smaller values of $\mathcal{L}/C_p$ to demonstrate the
clear agreement between analytical and numerical growth rates.

The simulations performed fall into three categories : (i) Linear simulations. These are non-dissipative runs performed with an intent to test the 
veracity of the linear analytical growth rate. (ii) Non-linear simulations. These fall into two groups (a) Short time high resolution runs to 
detect and quantify emergent coherent structures and their spectral scaling. (b) Long time lower resolutions runs to investigate the fate of 
differing sets of initial conditions, specifically those that have sufficient initial water substance and those that do not satisfy the 
sufficiency requirement. 

\subsection{Linear evolution}

\begin{figure}
\centering
\includegraphics[width=7cm,height=6cm]{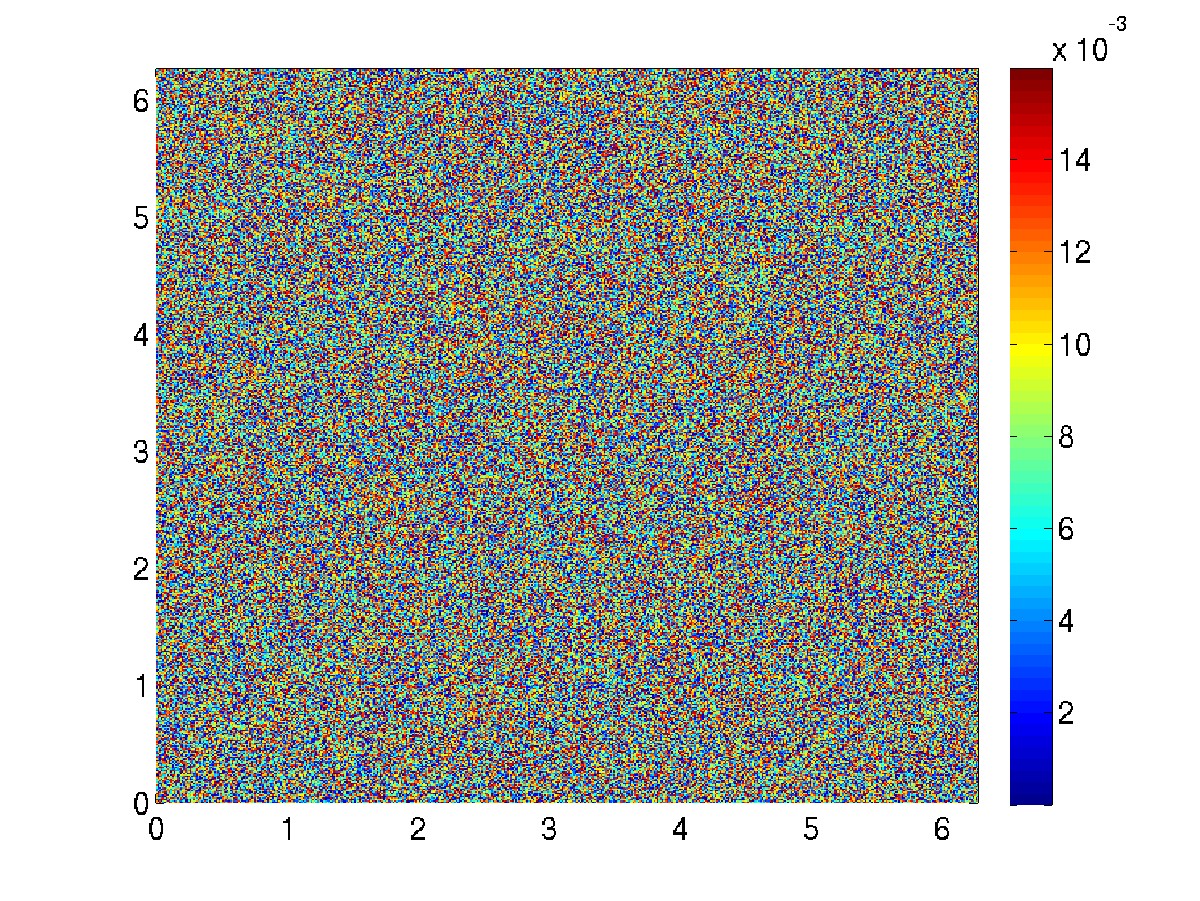}
\includegraphics[width=7cm,height=6cm]{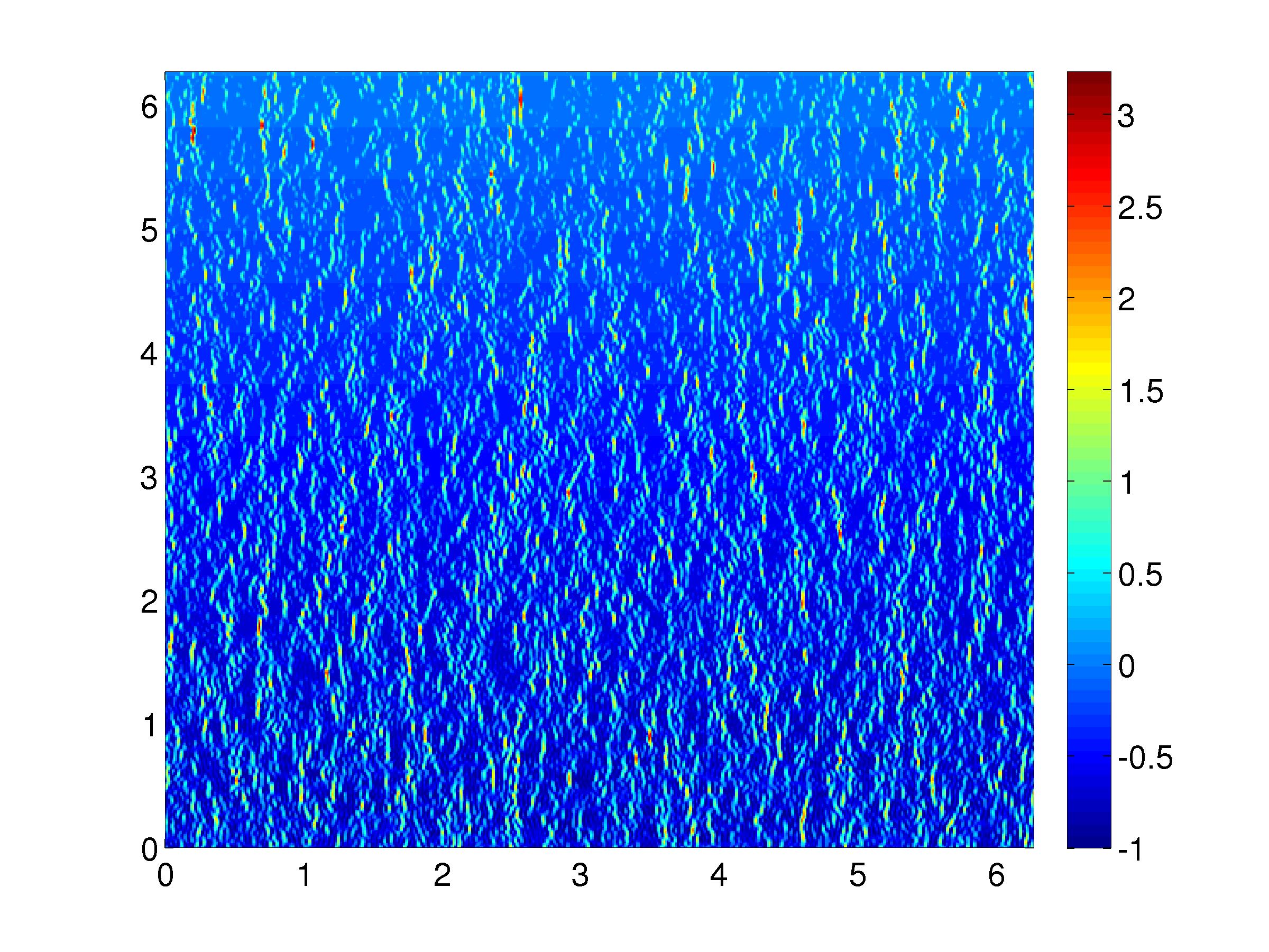}
\caption{\label{fig1} The first shows the scaled (i.e.\ $q_v'/Q_0$) spatially uncorrelated initial condition.
The second panel shows the emergent scaled perturbed vapor field for
$\tau_c=\tau_e=\epsilon=0.1$ at $t/\epsilon=10$ from the linear and non-dissipative simulation. }
\end{figure}

We begin with a set of linear non-dissipative simulations. 
Based on our rapid condensation ($\tau_c \rightarrow 0$) linear analysis, we expect vertically coherent 
structures to be the most unstable modes. As the maximal growth rate is independent of wavenumber, we examine the 
growth of spatially uncorrelated initial data with equal power at all scales. 
We choose $\tau_c=\tau_e=\epsilon=0.1$ and the resolution of the simulations is $512 \times 512$. 
Figure \ref{fig1} shows the initial 
condition and its fate after linear evolution for $10$ nondimensional units of time ($t/\tau_c$). 
Clearly, in spite of the ambient stratification, in agreement with our linear analysis the emergent vapor field exhibits a degree of vertical coherence. 

\begin{figure}
\centering
\includegraphics[width=9cm,height=7cm]{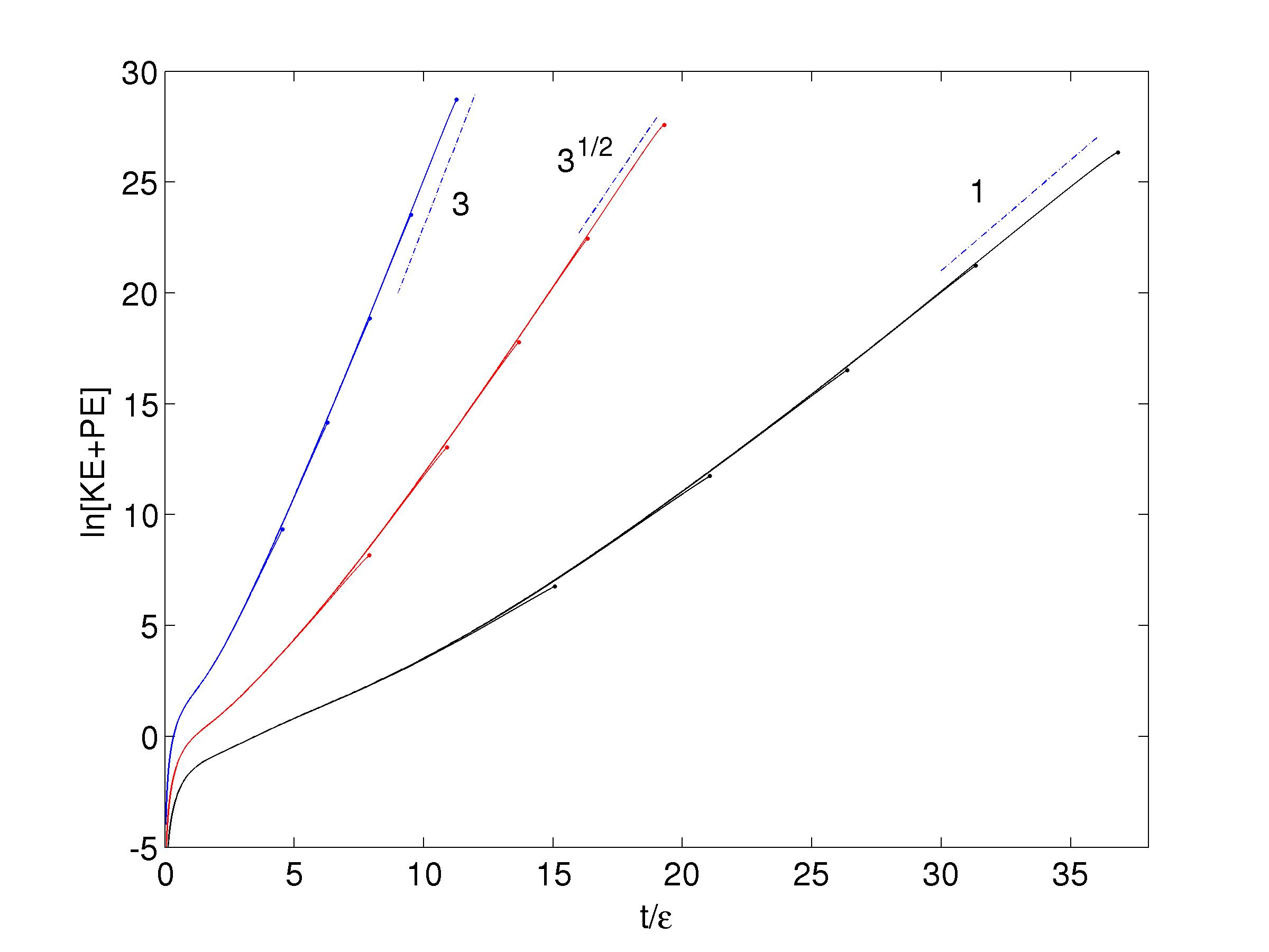}
\caption{\label{fig3} A plot of the $\log$ of the LHS of (\ref{1c}) Vs $t/\epsilon$ for single scale initial conditions. 
Smaller $\epsilon$ runs are carried out to progressively longer times, and for a fixed $\mathcal{L}/C_p$ the   
the curves collapse on top of
each other. The dashed lines show the analytical scaling of $\sqrt{\frac{\mathcal{L}}{C_p}-1}$ for $\mathcal{L}/C_p = 10,4$ and $2$ respectively. }
\end{figure}

The linear rapid condensation analysis, 
after substituting $g/T_0=\lambda=\beta=1/\epsilon$, yields a maximal growth rate of $\frac{1}{\epsilon}\sqrt{\frac{\mathcal{L}}{C_p}-1}$.
We proceed to compute the numerical growth rate of the total dry energy (LHS of equation \ref{1c}) for varying 
$\epsilon (=\tau_c=\tau_e)$ and $\mathcal{L}/C_p$. In particular, we let $\epsilon$ vary from $10^{-1}$ to $10^{-4}$ for $\mathcal{L}/C_p = 10,4$ and $2$ respectively.
As seen in Figure \ref{fig3}, on appropriate rescaling of the time (i.e.\ plotting vs $t/\epsilon$), all curves for a fixed 
$\mathcal{L}/C_p$ collapse on 
top of each other. Further, the numerical growth agrees with the analytical estimate, and this agreement improves as 
$\epsilon=\tau_c \rightarrow 0$, i.e.\ we approach the limit of rapid condensation. 

\subsection{Nonlinear evolution : moist turbulence}

We now proceed with nonlinear simulations to detect and quantify the emergence of scaling and coherence from initially spatially uncorrelated conditions. 
As before $g/T_0=\lambda=\beta=1/\epsilon$, and $\frac{\mathcal{L}}{C_p}=1/\epsilon$.
The simulations are carried out at a resolution of $750 \times 750$ and we begin with 
$\epsilon=0.1$. 

\begin{figure}
\centering
\includegraphics[width=7cm,height=6cm]{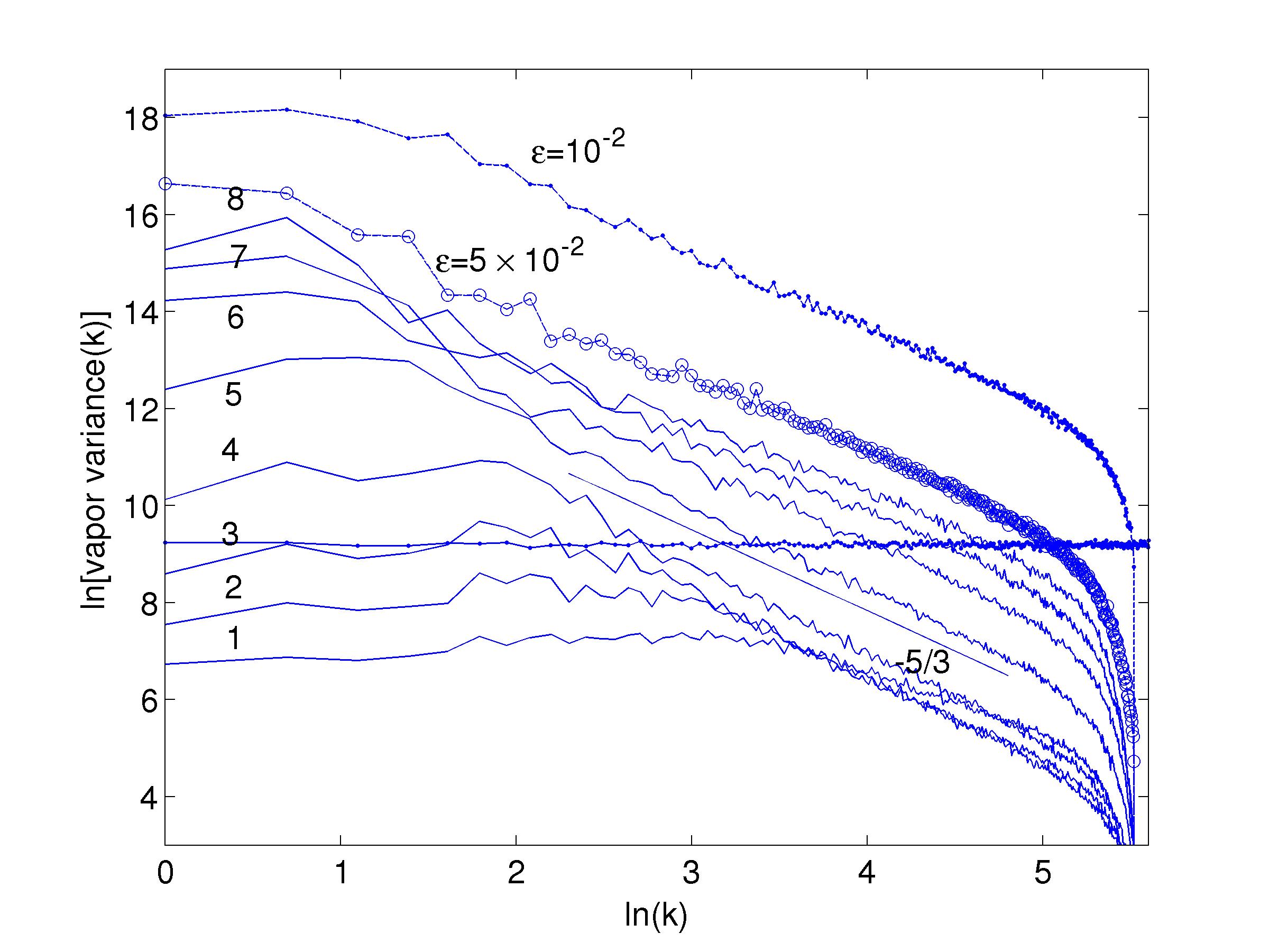}
\includegraphics[width=7cm,height=6cm]{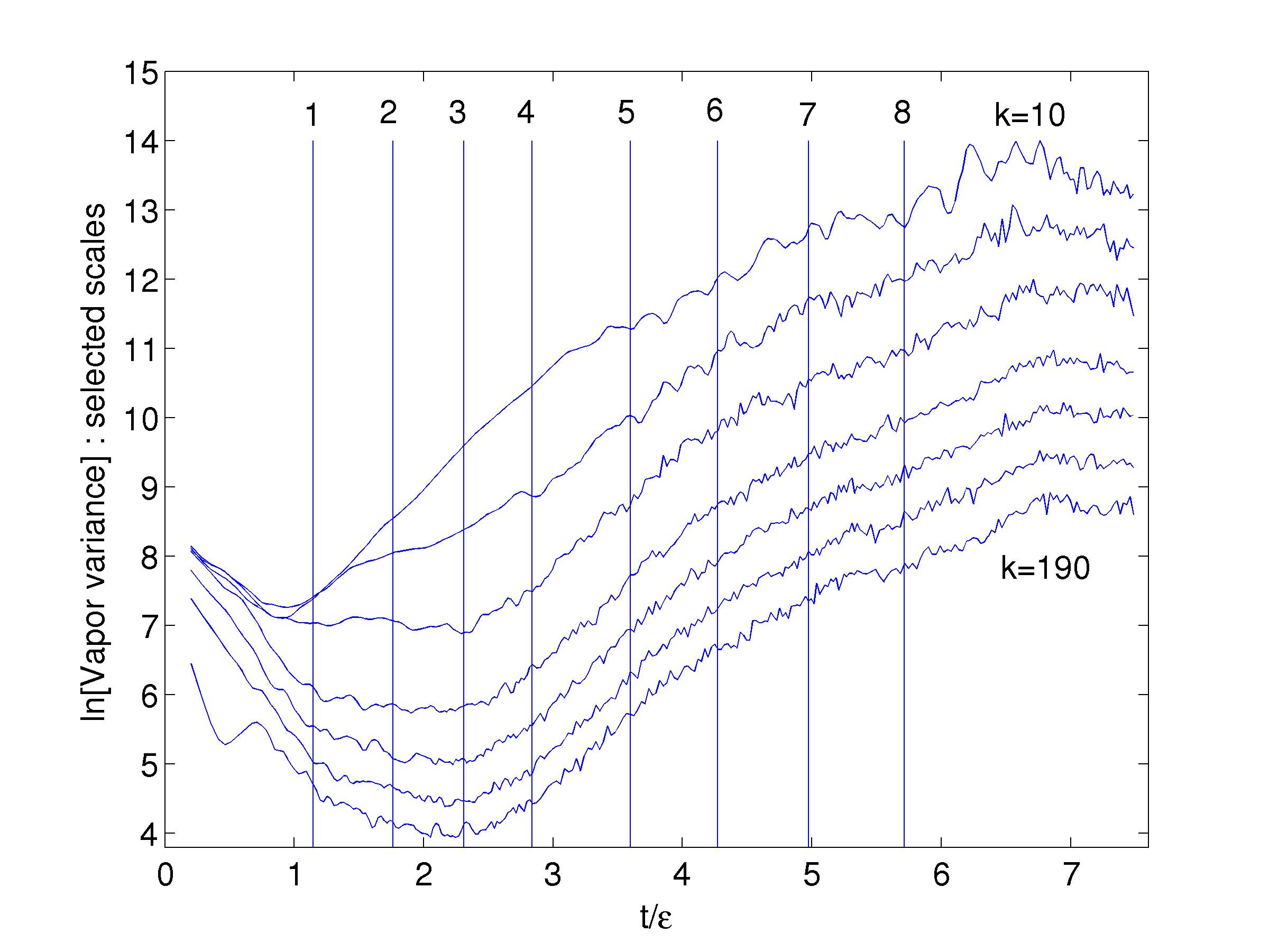}
\caption{\label{fig4} First panel : A plot of the $\log$ of the vapor variance for $\epsilon=0.1$ as it evolves in time. The numbers on the plots indicate the time ordering
(note that the spacing between these time slices is not equal). The flat line is the initial power spectrum, which as prescribed is uniform in
wavenumber. The uppermost two curves with dots and open circles show the invariant vapor variance spectrum 
for $\epsilon=5 \times 10^{-2}$ and $\epsilon=10^{-2}$ respectively.
Second panel : The growth of variance in time with $\epsilon=0.1$ for wavenumbers $k=10,20,40,80,120,160$ and $190$, ordered from top to bottom. The
vertical lines denote the times when the spectra are plotted in the first panel. }
\end{figure}

Figure \ref{fig4} shows the evolution of the power spectrum of the vapor 
variance (i.e.\ $(q_v')^2$). 
As the initial condition is super saturated, at small times we have condensation over the entire domain
which manifests itself in a drop in the total vapor variance. Subsequently, as observed via the sequentially numbered curves in the first 
panel of Figure \ref{fig4}, the spectrum evolves due a systematic 
growth of variance at all scales. 
As seen in the second panel of Figure \ref{fig4}, which shows the variance in selected wavenumbers as a function of time,
after the initial period of decay due to condensation throughout the domain, the growth of vapor variance is almost exponential at all scales.  
Though, the
growth begins earlier for progressively larger scales, hence in a short time the spectrum attains 
a characteristic negative slope.
Finally, the variance at all scales levels off and
the spectrum attains an invariant form (with fluctuations at the largest scales). 

To test the robustness of this behavior, we have 
repeated the simulations for smaller $\epsilon$, specifically for $\epsilon=5 \times 10^{-2}$ and $10^{-2}$. As mentioned, with decreasing $\epsilon$, we have  
stronger stratification and we also approach the limit of rapid condensation/evaporation. The invariant form of the vapor variance 
spectrum in
these two cases is shown in the first panel of Figure \ref{fig4}. Clearly, the spectrum remains the same for the range of $\epsilon$
considered. The growth of the vapor variance in different scales (not shown) also follows a similar pattern to that in the second panel of
Figure \ref{fig4}.

A similar growth of energy at all scales is also observed (not shown) in the velocity and temperature fields, though the invariant spectrum 
obtained follows different power laws. With regard to the kinetic energy we obtain a steeper spectrum (a power law with approximately -2 slope) while the 
temperature variance or potential energy exhibits shallower scaling (a power law with approximately -1 slope). Once again, these spectra 
show the same scaling for the range of $\epsilon$ we have considered.

Even though the various spectra show an evolution that is reminiscent of the inverse cascade in 2D turbulence (though
with a remarkable diversity of power laws for the various fields), the mechanism of 
generation of this spectrum is quite different from the interscale transfer that characterizes 2D flows. Specifically, in the present 
unstable setting, the vapor, velocity and temperature fields experience a growth of energy at all scales and the spectral form 
emerges due to the comparatively larger growth at large scales.

\begin{figure}
\centering
\includegraphics[width=5cm,height=4.5cm]{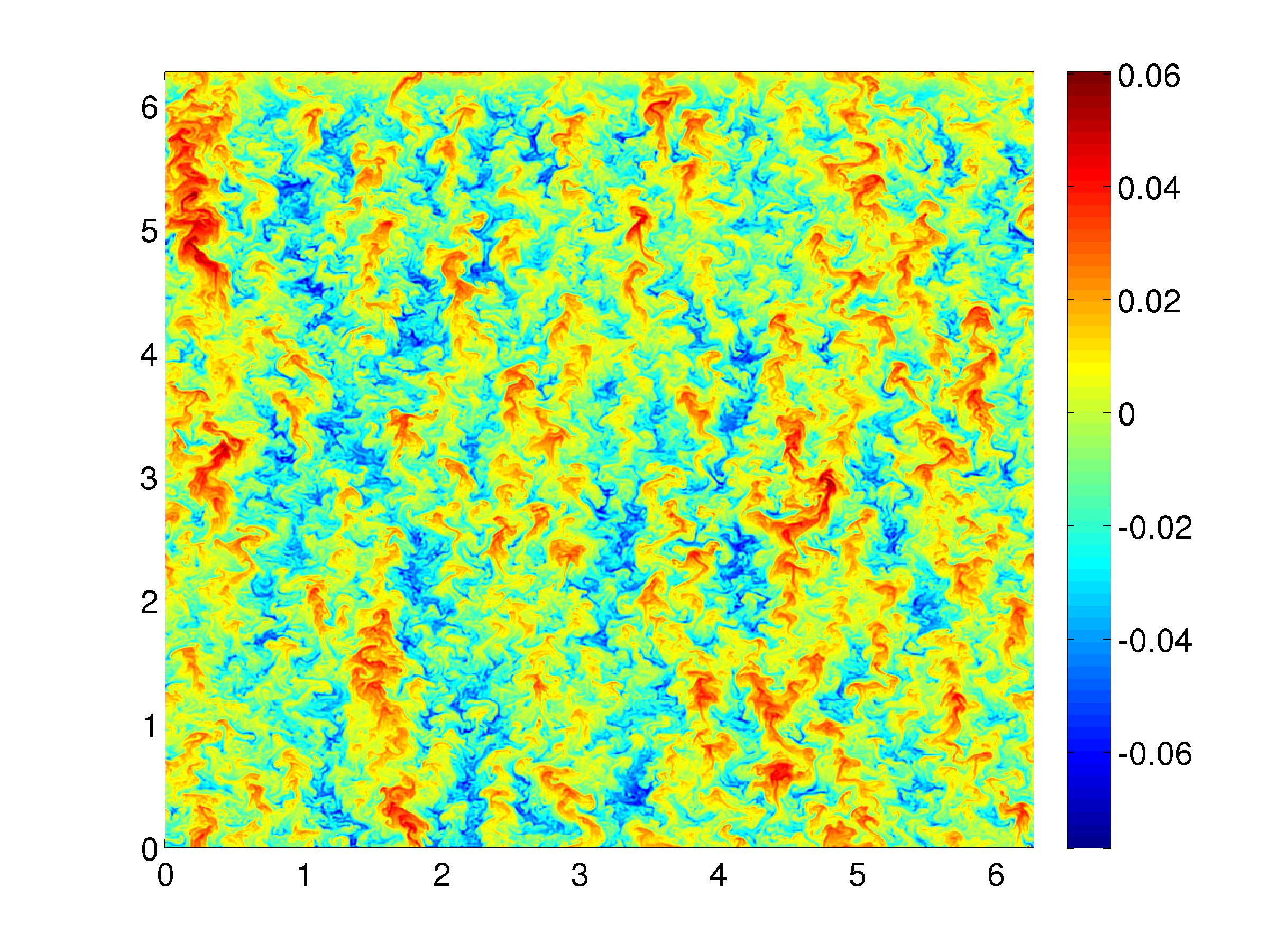}
\includegraphics[width=5cm,height=4.5cm]{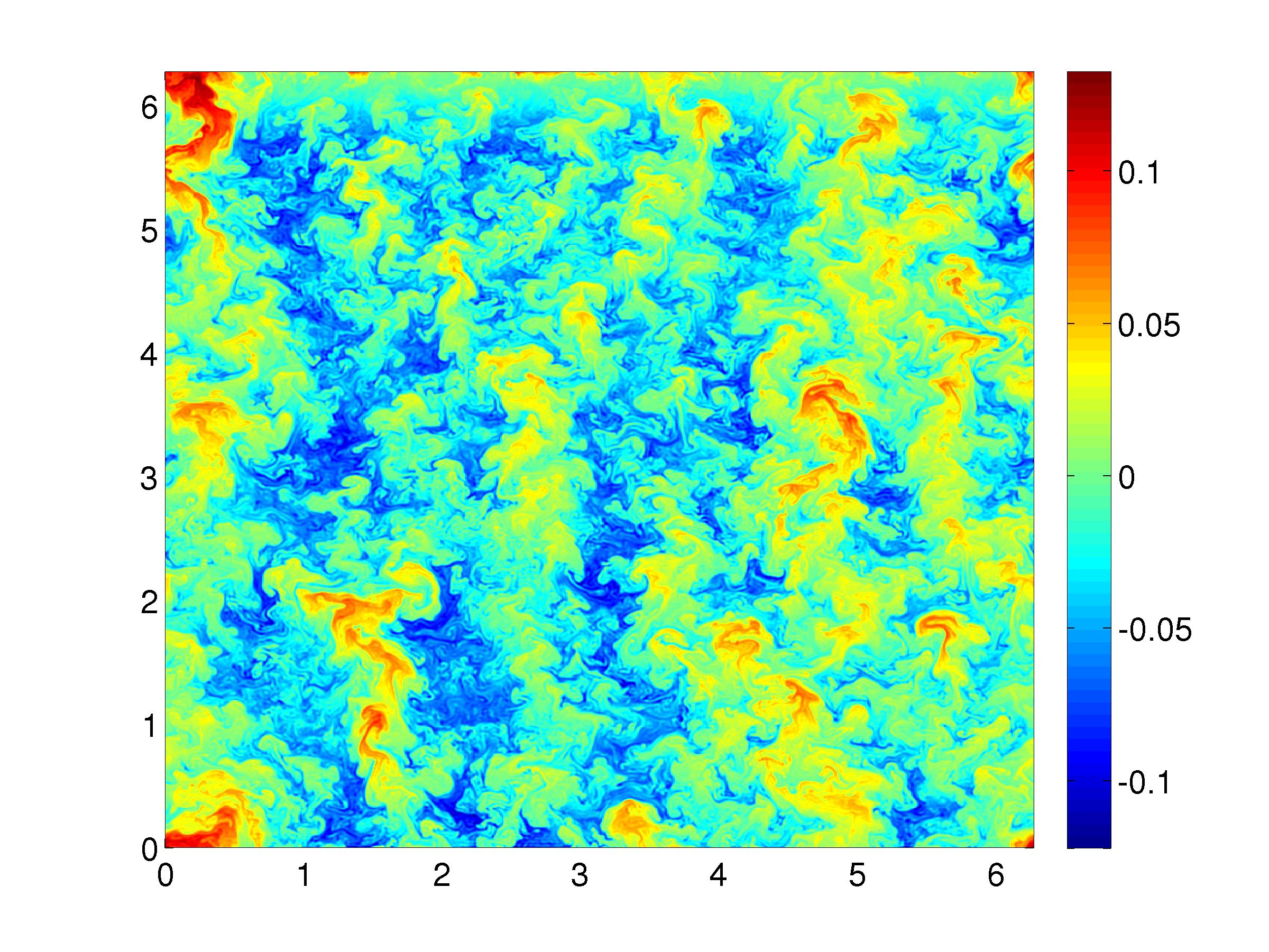}
\includegraphics[width=5cm,height=4.5cm]{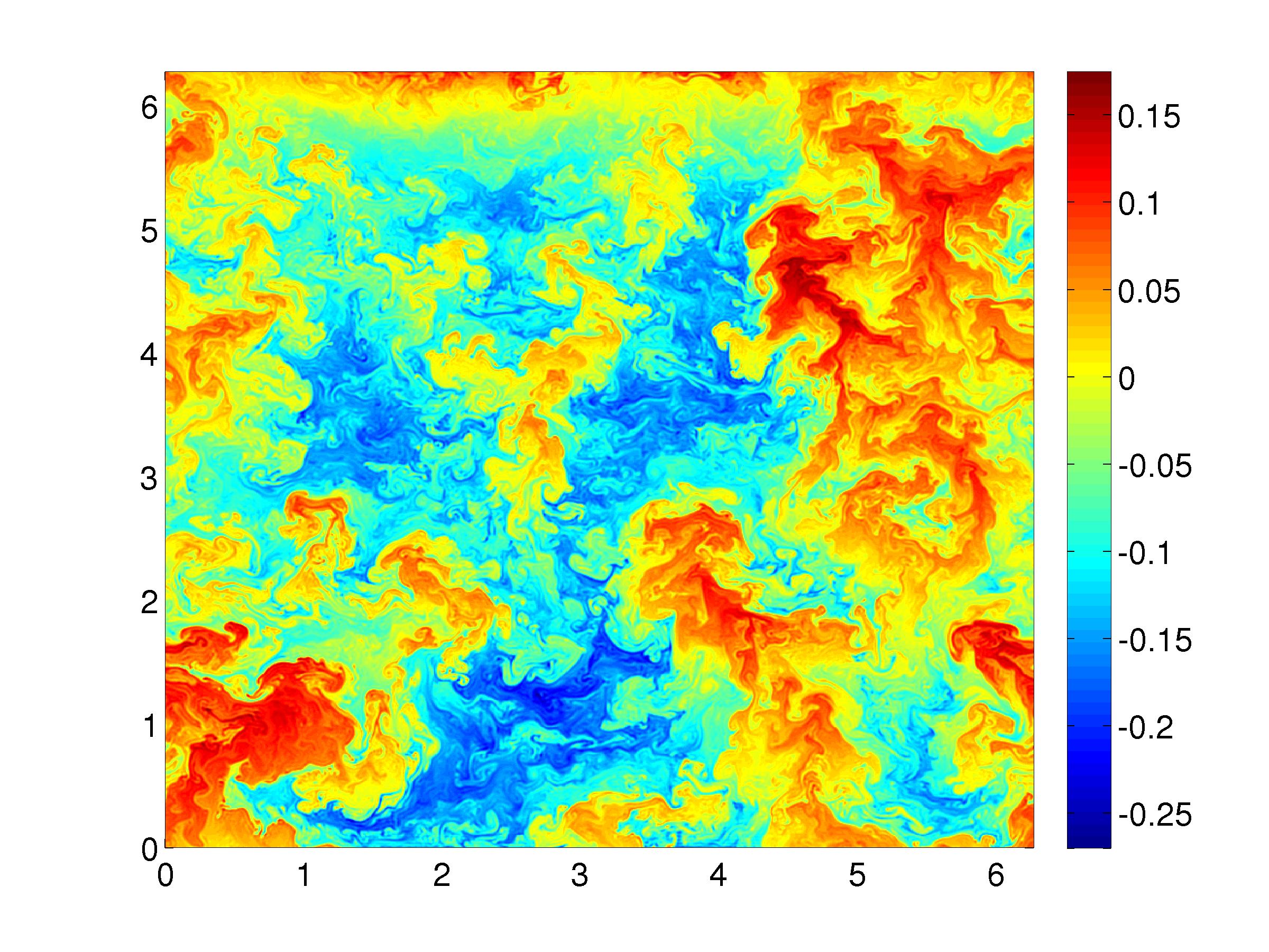}
\caption{\label{fig5} Evolution of a spatially uncorrelated initial condition for $\epsilon=10^{-1}$.
The three panels show the emergent perturbed vapor field ($q_v'/Q_0$) for
the time labels 4, 5 and 7 as referred to in Figure \ref{fig4}. The initial condition can be seen in the first panel of Figure \ref{fig1}. }
\end{figure}

Along with the spectral evolution of the various fields, we also notice the 
emergence of ordered structures 
from the initially spatially uncorrelated 
data. An example of this is can be seen in Figure \ref{fig5} which, for $\epsilon=0.1$, shows the perturbed vapor field in various stages of its evolution. 
We refer to this state, i.e.\ one with multiscale coherent structures, as a state of moist turbulence. In the following sections, we inquire into the 
long term fate of moist turbulence for differing classes of initial conditions.

\subsection{Nonlinear evolution : long time behavior}

We now proceed to the long-time nonlinear evolution of the system. 
The simulations are carried out at a resolution of $350 \times 350$ for
$\epsilon=10^{-1}, 5 \times 10^{-2}$ and $10^{-2}$. 
In all cases, the evolution was seen to be similar. We focus on the case $\epsilon=10^{-2}$, but comparative plots from other runs are
also presented. 

\subsubsection{Initial data with sufficient water substance}

We first consider initial data such that the water substance is sufficient to saturate the entire domain. 
To satisfy this condition we prescribe 
supersaturation everywhere, and as before $q_v'$ is a uniformly distributed spatially uncorrelated field.
As expected the vapor field evolves from its initially uncorrelated state, and coherent structures
develop along with a systematic spectral evolution. In fact, the early behavior of the system and the development of moist turbulence is described in
Section III B. 

\begin{figure}
\centering
\includegraphics[width=7cm,height=6cm]{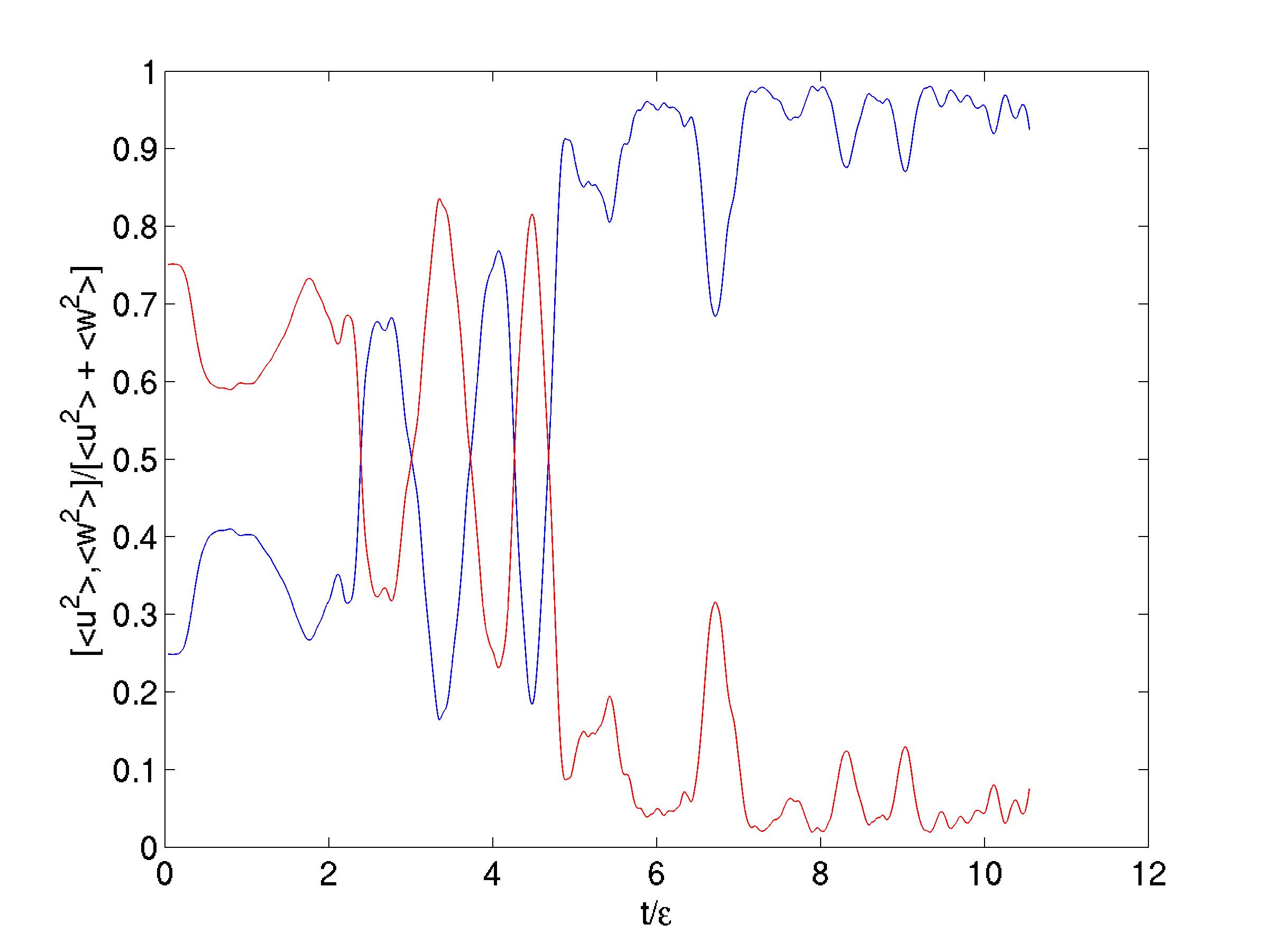}
\includegraphics[width=7cm,height=6cm]{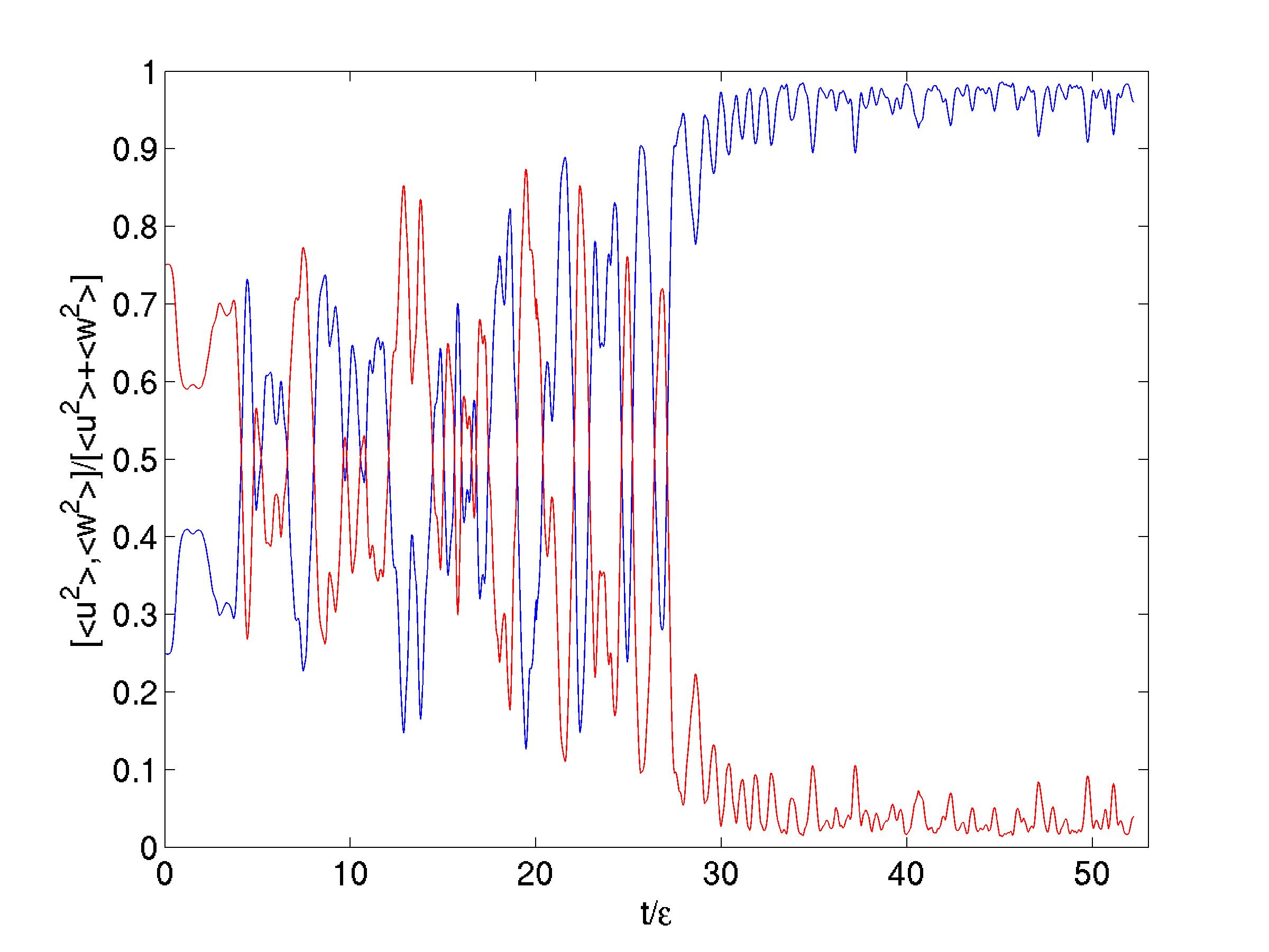}
\caption{\label{fig8} The first two panels show the contribution of the horizontal (blue) and vertical (red) components to the kinetic energy of the flow as a function of
the nondimensional time ($\frac{t}{\epsilon}$) for $\epsilon=10^{-2}$ (first panel) and $\epsilon=5 \times 10^{-2}$ (second panel).} 
\end{figure}

\begin{figure}
\centering
\includegraphics[width=7cm,height=6cm]{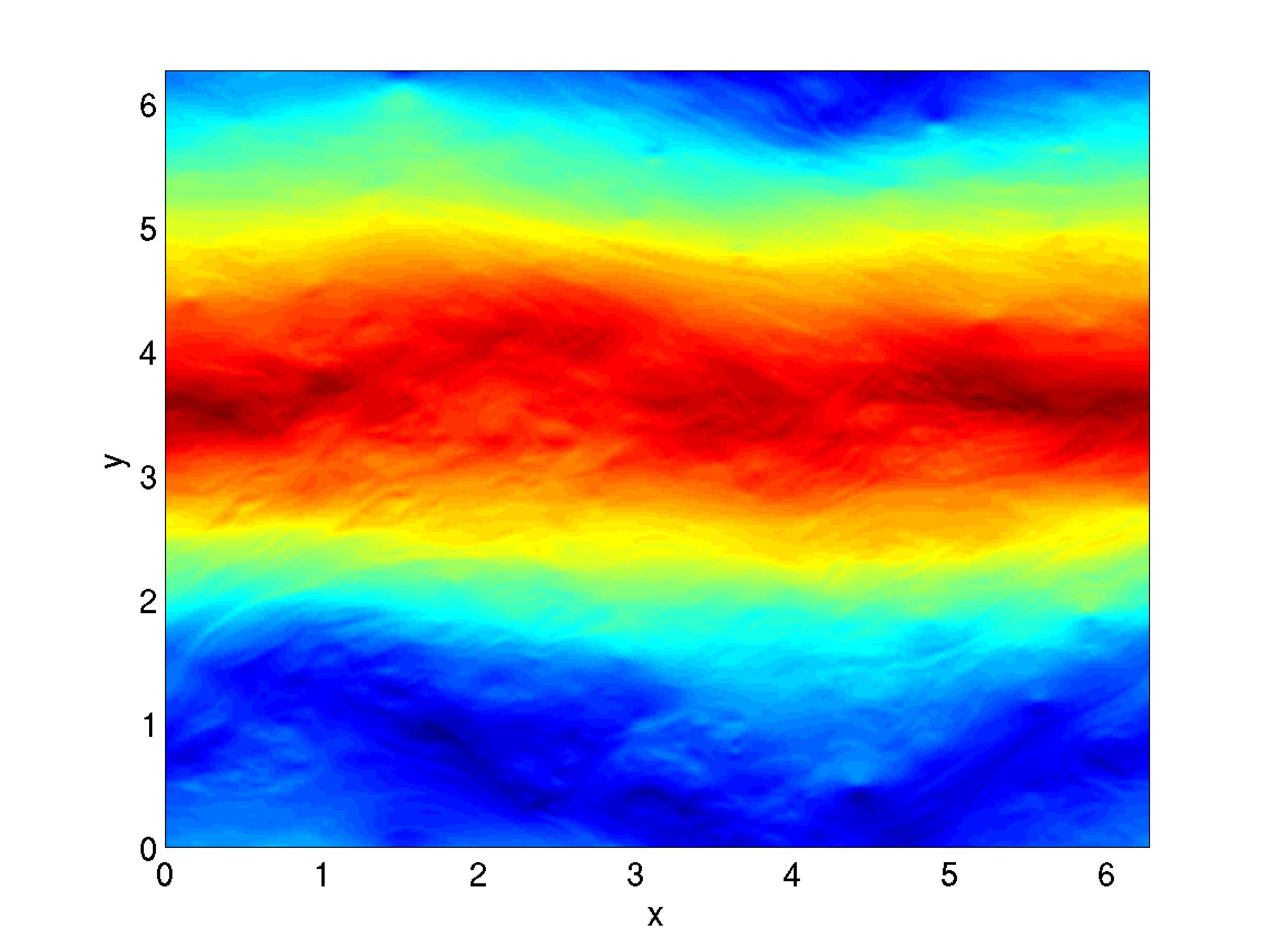}
\caption{\label{fig11} The horizontal flow with sufficient initial water substance for $\epsilon=10^{-2}$ at $\frac{t}{\epsilon}=10$.} 
\end{figure}

After the spectra of the various fields attain an invariant form, we follow the evolution of the system. In particular, the first two panels of Figure \ref{fig8} show the horizontal
and vertical components of the kinetic energy as functions of nondimensional time for all $\epsilon=10^{-2}$ and $5 \times 10^{-2}$. In both cases, 
at early times the components oscillate with 
comparable contributions to the kinetic energy from the vertical and horizontal portions. But, at a critical time (${\tau_\epsilon}^*$) the system experiences a rather 
abrupt transition beyond which the horizontal component dominates the kinetic energy budget. Note that ${\tau_\epsilon}^*$ decreases with $\epsilon$, i.e.\ 
the transition to horizontal flow dominance is progressively more rapid as $\epsilon \rightarrow 0$.
Focussing on $\epsilon=10^{-2}$, Figure \ref{fig11} shows the 
horizontal flow after $\frac{t}{\epsilon}={\tau_\epsilon}^* \approx 5$ : as is evident, we observe the 
formation of a VSHF. The VSHF in each case is broad and the flow in each direction occupies approximately half the domain.
Though not shown, the development of a VSHF is accompanied by a tendency to saturation. Further, as the vertical velocity becomes very weak, the system 
also experiences a progressively diminishing amount of heating from condensation and evaporation. In fact, as per (\ref{6f}), the liquid water and temperature
are passively advected by the VSHF.

Gathering these pieces together we argue that subsequent to the development of a moist turbulent state, at long times, with sufficient initial water 
substance, in accordance with the asymptotic state of moist balance, the strongly stratified and rapidly condensing/evaporating system approaches
a saturated VSHF. 

\subsubsection{Initial data with insufficient water substance}

\begin{figure}
\centering
\includegraphics[width=7cm,height=6cm]{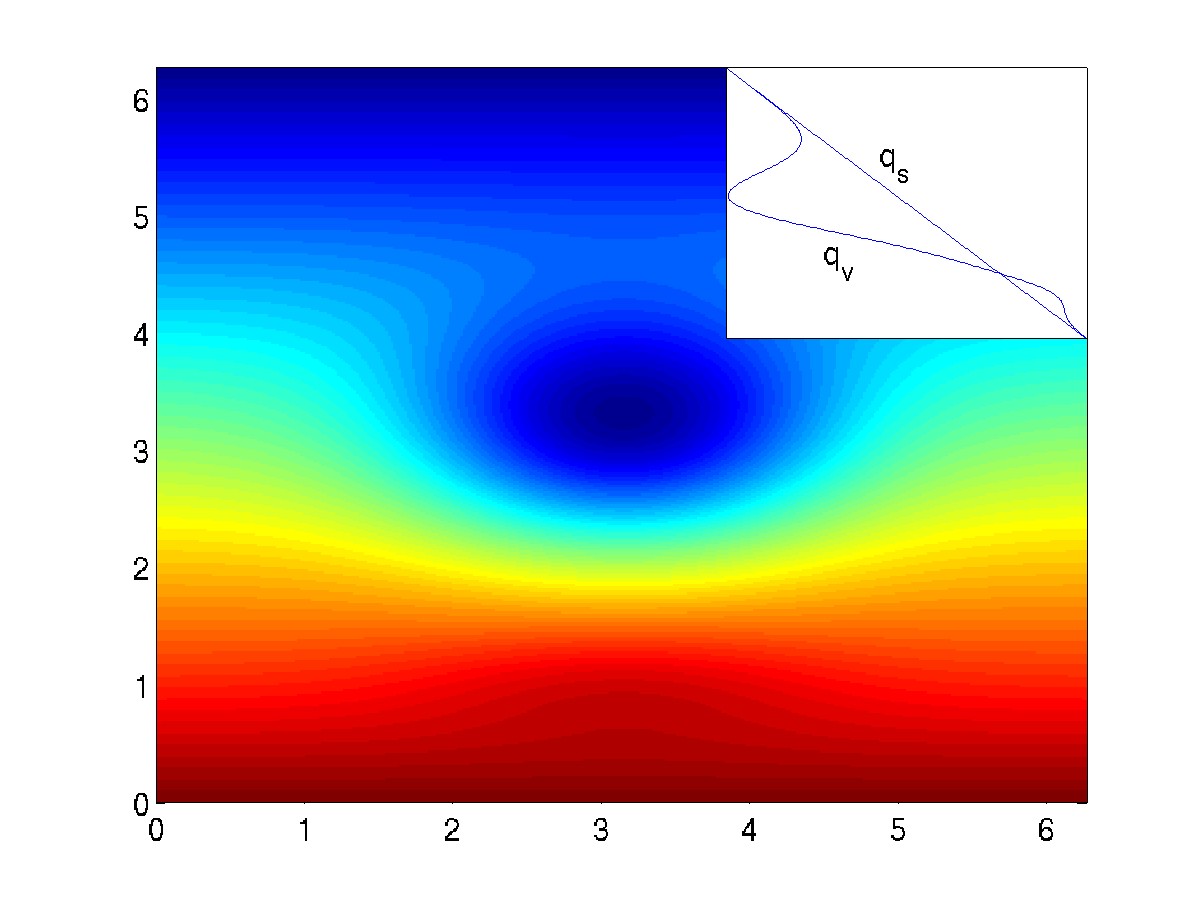}
\caption{\label{fig11a} 
The initial vapor field when the initial water substance cannot saturate the domain, the inset shows a cross-section of $q_v$ and $q_s$
through the center of the domain.}
\end{figure}

We now proceed to initial data where the initial water substance cannot saturate the domain. 
As these are very long time simulations, they are carried out at a resolution of 256 $\times$ 256 grid points.
To satisfy this condition we supersaturate the domain in a localized Gaussian blob, while undersaturating 
it in a similar manner in the rest of
the domain. A snapshot of the initial vapor field is shown in Figure \ref{fig11a}. 
We ensure that $q_v(x,z,t=0) \ge 0$, the total initial water substance is less than the amount required for the domain to be saturated
and that ${q_v}'$ is a spatially periodic
function. The region of supersaturation ensures that the system evolves without external forcing. 

\begin{figure}
\centering
\includegraphics[width=7cm,height=6cm]{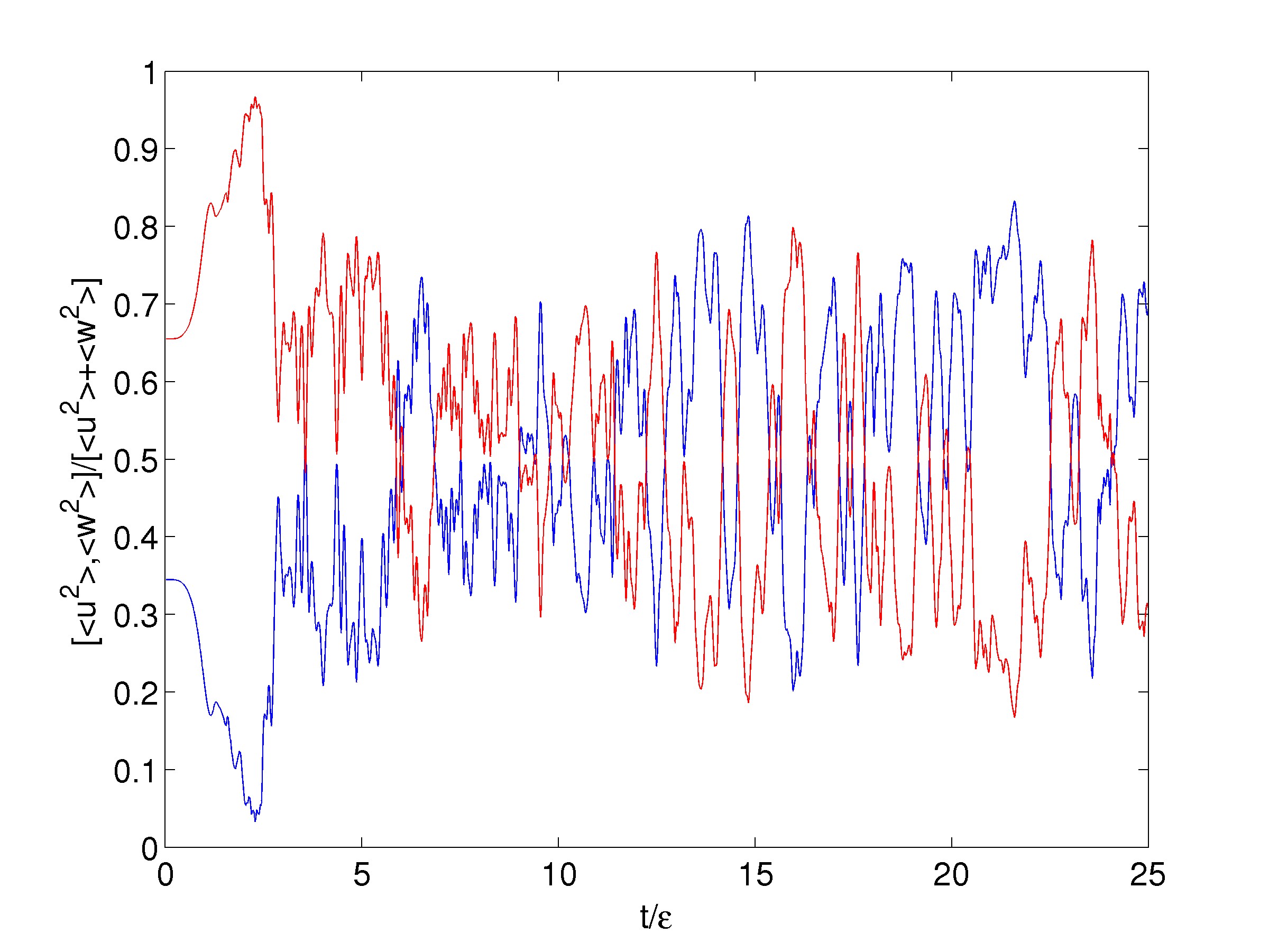}
\includegraphics[width=7cm,height=6cm]{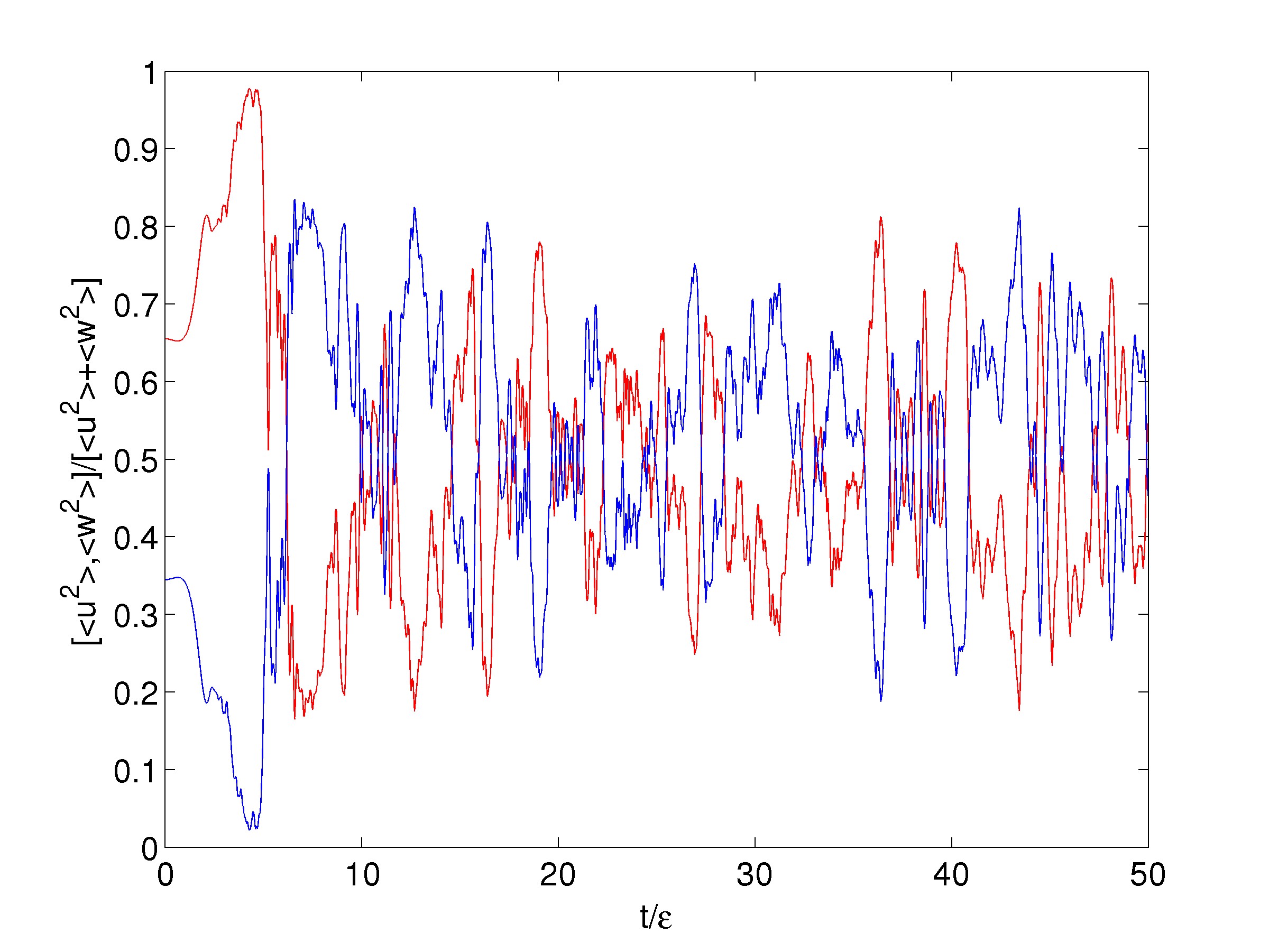}
\caption{\label{fig12} Same as Figure \ref{fig8} but with insufficient initial water substance as shown in the second panel of Figure \ref{fig11}. The simulations have been 
carried out for a significantly longer duration than shown in these plots.}
\end{figure}

As with the previously considered sufficient class of initial data, the spectral scaling of the various fields evolves to 
the invariant state described in Section IIIB. After the development of a turbulent flow, we once again follow the evolution of the system.
In particular, as in Figure \ref{fig8}, the two panels of Figure \ref{fig12} 
show the vertical and horizontal contribution to the kinetic energy as a function of
nondimensional time for $\epsilon=10^{-2}$ and $\epsilon=5 \times 10^{-2}$. In contrast to initial data with sufficient water substance,
clearly we do not 
observe a transition to horizontal flow dominance. Further, as expected, the domain does not tend to saturation and there is persistent heating/cooling 
throughout the domain.

\begin{figure}
\centering
\includegraphics[width=9cm,height=7cm]{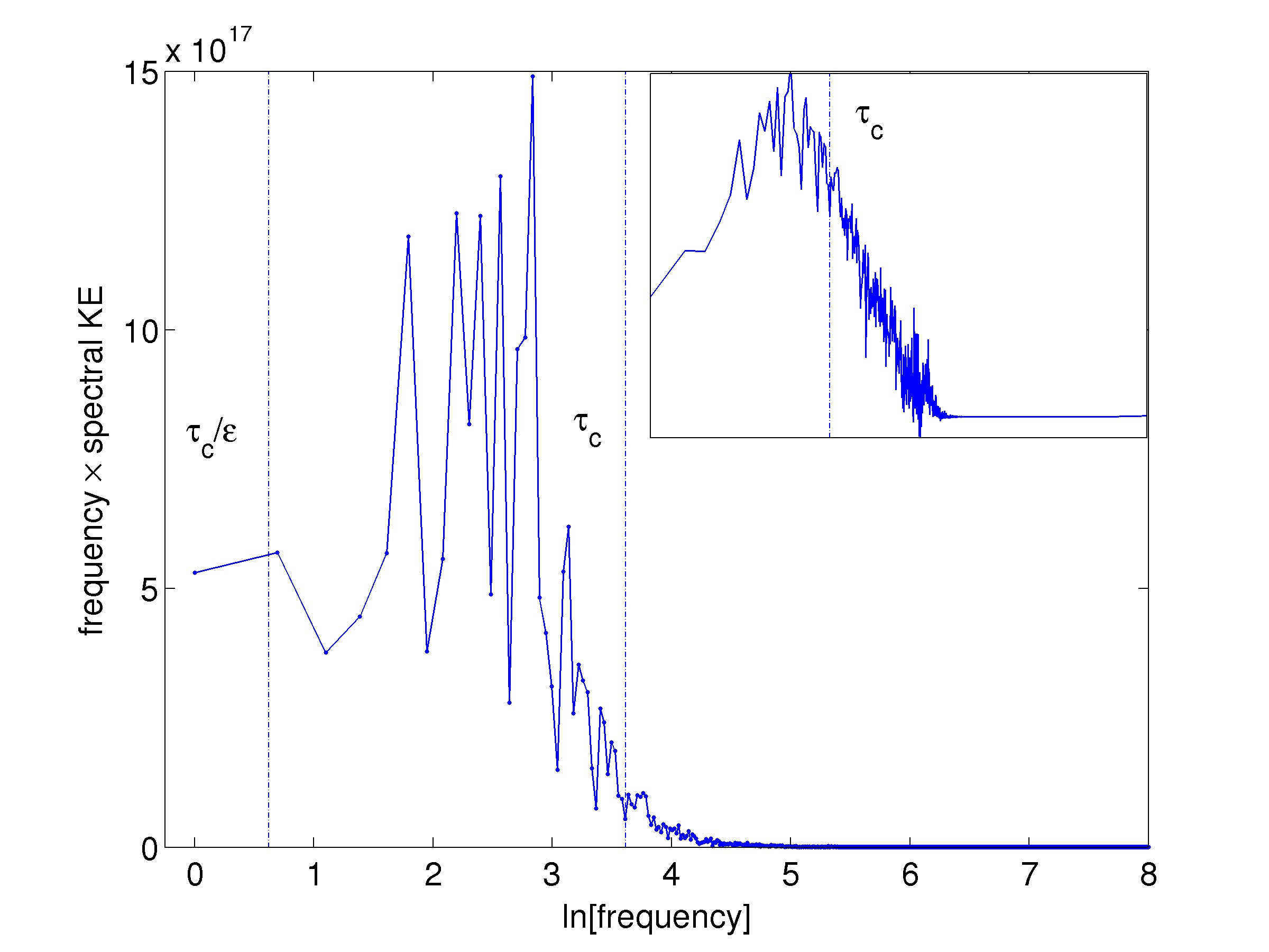}
\caption{\label{fig13} The power spectrum of the kinetic energy component time series seen in Fig.\ (\ref{fig12}) for $\epsilon = 5 \times 10^{-2}$. 
We consider an average over twelve long segments,
each encompassing multiple oscillations seen in the signal. The main plot shows the variance preserving form of the spectrum. The inset shows the red compensated 
spectrum, i.e.\ spectrum after a multiplication by the square of the frequency.}
\end{figure}

Examining Figure \ref{fig12}, we note that the large amplitude oscillations are much slower than the imposed scale of condensation/evaporation (this is 
evident as the time in the plots has been nondimensionalized by $\tau_c$). 
Figure \ref{fig13} shows the
variance preserving temporal spectrum of the horizontal and vertical components of the kinetic energy for $\epsilon=5 \times 10^{-2}$. 
As expected, there is a large amount of power at small frequencies or
large time scales. Further, we can clearly see multiple peaks that lie between $\tau_c$ and $\tau_c/\epsilon$ which indicates a slow (as compared to $\tau_c$) aperiodicity in the time 
series. 
Also, at small time scales (or large frequencies), especially smaller than $\tau_c$, the flattening of the red-compensated spectrum shown in the inset of Figure \ref{fig13} 
indicates a red noise scaling.

The broad nature of the peak in the inset of Figure \ref{fig13}, as well as the multiple peaks seen in the variance preserving form of the 
spectrum, suggests an aperiodic or quasiperiodic character to the slow oscillations seen in the 
components of the kinetic energy. Performing a spectral analysis of the domain averaged heating 
as well as the mean vapor and liquid water content (not shown) reveals a similar broad peak at long time scales 
compared to $\tau_c$. Quantitatively, the slow time scale has a range from double to approximately $1/\epsilon$ times the 
imposed condensation timescale. We have 
verified that simulations with other values of $\epsilon$ (specifically, $\epsilon=10^{-1}$ and $10^{-2}$) yield
similar results.

Therefore, as expected from the asymptotic analysis, when the initial conditions do not have enough water substance to saturate the domain, 
the system does not attain a state of moist balance. In fact, the system oscillates about moist balance. This can be understood from Figure \ref{fig12} where we 
see that, as the contribution of the horizontal flow kinetic energy increases, the system approaches moist balance. But, as the entire domain can never be saturated, the 
vertical flow kinetic energy soon gains prominence. This cycle repeats aperiodically at the
aforementioned slow time scale. It is important to note that while the domain averaged characteristics are modulated at a slow time scale, the various fields 
maintain their spectral scaling and the system remains in a state of moist turbulence. Finally, even though it is not possible to associate a single
timescale with the slow oscillation, we see that its broad peak extends from twice to $1/\epsilon$ times the imposed scale of condensation 
(and evaporation).

\section{Conclusion and discussion}

We have considered a 2D moist Boussinesq system in a periodic setting with simple condensation and evaporation schemes. In particular, these
schemes are designed to push individual parcels towards a saturated state. This allows possibly the simplest scenario 
wherein we can study the interaction 
of advection with evaporation and condensation in a dynamically consistent manner. Linearizing the problem about a state of saturation, with no 
liquid water and hydrostatic balance showed that the most unstable modes were vertically coherent. In fact, in the limit of rapid 
condensation, the form of the growth rate of the unstable modes took a particularly simple form. 
A suite of 
non-dissipative linear 
simulations verified these predictions. 

One of the main questions we wished to address in this work was the development of initially spatially uncorrelated vapor fields. Keeping in mind that the 
2D Boussinesq system does not conserve enstrophy, and that there is significant theoretical and numerical evidence showing that the system 
supports a transfer of energy from 
large to small scales \cite{McEwan-Rob},\cite{Mied},\cite{Klos},\cite{Orlanski},\cite{BS},\cite{BS1},\cite{SS}, 
it was not a priori evident to us that initially uncorrelated fields 
would develop into coherent structures, or that 
we would see a growth of energy at large scales. Pursuing nonlinear simulations we observed that coherent structures do emerge spontaneously from 
spatially uncorrelated initial conditions. Further, the scaling of the vapor variance as well the kinetic and potential energy attained an 
invariant form and follows a power-law. In fact, the system attains a state that can be aptly described as moist turbulence.

Interestingly, during the nonlinear simulations, the growth of energy at large scales cannot be attributed to an inverse transfer kind of mechanism 
seen in pure 2D turbulence wherein energy is transferred from small to large scales. In fact, in an unstable setting, 
the variance in all fields grew
very rapidly at all scales and this rapid growth persisted for a
longer time at larger scales leading to a characteristic negatively sloped spectrum. Given that the emergence of coherent structures in the vapor field 
did not depend on the conservation of enstrophy, we are optimistic that a similar development of multiscale structures 
from initially uncorrelated fields 
will be possible in a three-dimensional 
moist stratified system. 

Our other goal was to examine the notion of balance and slowness in physically relevant limits of this idealized moist system. 
Imposing strong stratification (implying the 
presence of high frequency gravity waves) along with rapid condensation (implying rapid heating/cooling via condensation/evaporation), 
we found that, if there is sufficient initial water substance to saturate the entire domain, then the system supports a hydrostatic and saturated 
VSHF (a state of moist balance). If the initial water substance is not sufficient to saturate the domain, then the state of moist balance is inaccessible and the system evolves 
via a continual interaction of evaporation and condensation in different regions of the domain.

We began our numerical investigation of the long time development of the system by considering initial conditions 
that have sufficient water substance.
Simulations were carried for for a suite of $\epsilon$ values (for progressively stronger stratification and smaller condensation timescales), 
and the evolution in each case was similar. As per
expectations the system evolved with coherent structures and scaling emerging from initially uncorrelated conditions. After the development of 
invariant spectra, we followed
the horizontal and vertical components of the kinetic energy as a function of nondimensional time ($t/\epsilon$). At 
small times, we noted oscillations with equal contribution from either 
component at different times. However, at $\frac{t}{\epsilon}={\tau_\epsilon}^*$, we observed a rather abrupt transition beyond which the horizontal 
component dominated the kinetic energy budget. Examining the state of the flow after $\frac{t}{\epsilon}={\tau_\epsilon}^*$,
we noted the formation of a persistent and dominant VSHF. In addition, the vapor field tended to saturation for $\frac{t}{\epsilon} \gg {\tau_\epsilon}^*$.
In essence, with sufficient initial water substance, we saw that, the system (at long times) tends to the
asymptotic state of moist balance. 

We then proceeded to initial conditions which were insufficient in initial water substance.
The now generic behavior involving coherent structures and scaling was observed, but as expected from the asymptotic analysis, the system did not evolve to a 
state of moist balance.
In fact, an analysis of the vertical and horizontal kinetic energy (as well as other domain averaged time series such as the heating and potential energy), 
revealed dominant slow aperiodic or quasiperiodic oscillations and a fine temporal structure 
within each such oscillation.
The period associated with these slow oscillations was much larger than the imposed condensation/evaporation time scale ($\tau_c$), and therefore the system 
was modulated on a relatively slow time scale. In particular, the slow oscillation was seen to range from twice to $1/\epsilon$ times the 
imposed timescale of condensation (and evaporation). We emphasize that the spectral scaling of the various fields (noted earlier) persists in these 
conditions, i.e.\ even though the system experiences a series of slow oscillations about the state of moist balance, all fields maintain their characteristic scaling and, in essence, the
system remains in a state of moist turbulence.

\vskip 1truecm
{\it Acknowledgements :}
JS would like to acknowledge many fruitful discussions with V.\ Venugopal (CAOS, IISc).
Financial support 
for JS and LMS was provided by NSF
CMG 0529596 and
the DOE Multiscale Mathematics program (DE-FG02-05ER25703).

\end{document}